\documentclass[twocolumn,prl,reprint,letterpaper]{revtex4-1}

\usepackage{changepage}

\usepackage{nameref}

\usepackage{lmodern}

\usepackage{graphicx}
\usepackage{dcolumn}
\usepackage{bm}
\usepackage{amsmath}
\usepackage{amssymb}
\usepackage{amsfonts}
\usepackage{color}
\usepackage{psfrag}
\usepackage{epstopdf}
\usepackage{natbib}
\usepackage{appendix}

\begin{document}

\title{Optimal Chemotactic Responses in Stochastic Environments}

\author{Martin God\'{a}ny$^{1,2}$}

\author{Bhavin S. Khatri$^1$}

\affiliation{%
$^1$MRC National Institute for Medical Research\\
Mathematical Biology Division\\
The Ridgeway, London, NW7 1AA, U.K.
}%
\author{Richard A. Goldstein$^2$}%
\email{r.goldstein@ucl.ac.uk}

\affiliation{%
$^2$Division of Infection and Immunity\\
University College London\\
London, WC1E 6BT, U.K.
}%

\begin{abstract}
Most of our understanding of bacterial chemotaxis comes from studies of \textit{Escherichia coli}. However, recent evidence suggests significant departures from the \textit{E. coli} paradigm in other bacterial species. This variation may stem from different species inhabiting distinct environments and thus adapting to specific environmental pressures. In particular, these complex and dynamic environments may be poorly represented by standard experimental and theoretical models. In this work, we study the performance of various chemotactic strategies under a range of stochastic time- and space-varying attractant distributions \textit{in silico}. We describe a novel type of response in which the bacterium tumbles more when attractant concentration is increasing, in contrast to the response of \textit{E. coli}, and demonstrate how this response explains the behavior of aerobically-grown \textit{Rhodobacter sphaeroides}. In this ``speculator'' response, bacteria compare the current attractant concentration to the long-term average. By tumbling persistently when the current concentration is higher than the average, bacteria maintain their position in regions of high attractant concentration. If the current concentration is lower than the average, or is declining, bacteria swim away in search of more favorable conditions. When the attractant distribution is spatially complex but slowly-changing, this response is as effective as that of \textit{E. coli}.
\end{abstract}

\maketitle



\section*{Introduction}
Movement of \textit{Escherichia coli} consists of periods of running punctuated by tumbling events where the bacterium randomly changes direction. This can result in successful chemotaxis when the probability of initiating a tumble per short time interval (the tumbling rate) is a function of the concentration of attractant experienced by the bacterium. In the absence of attractant, an \textit{E. coli} bacterium has a constant, basal tumbling rate. When the attractant concentration experienced by the bacterium is increasing, tumbling events become less frequent, so the bacterium has longer runs in the direction of increasing attractant. Conversely, when the bacterium detects a decreasing attractant concentration, it tumbles more often, shortening its runs down attractant concentration gradients. In \textit{E. coli}, this mechanism involves an excitation pathway, inhibiting tumbling, and an adaptation pathway that methylates the receptors, decreasing their sensitivity and thereby attenuating the excitation pathway in the continued presence of attractant. This ``adaptive'' response allows the bacterium to locate and stay in regions of high attractant concentration. When attractant concentration plateaus, the tumbling rate returns to the basal rate, a phenomenon called ``perfect adaptation''. A consequence of perfect adaptation is that the response is independent of the absolute concentration of attractant and depends only on differences in concentration experienced by the bacterium.

While \textit{E. coli} has been instrumental for our understanding of chemotaxis, other bacteria show a considerable variety of chemotactic mechanisms and behaviors~\cite{Wuichet2010, Hamer2010}. For example, responses have been identified which show very little adaptation, such as in certain cultures of \textit{Rhodobacter sphaeroides}~\cite{Poole1988}. Furthermore, aerotaxis in \textit{E. coli} is thought to involve the Aer receptor which lacks methylation sites~\cite{Armitage1997}, suggesting a lack of adaptation. Strangely, some bacteria seem to tumble more in the presence of attractant. This appears to be the case in many mutant strains, such as aerotaxis~\cite{Dang1986} and redox taxis~\cite{Bespalov1996} in mutated \textit{E. coli}, aerotaxis in mutated \textit{Salmonella typhimurium}~\cite{Dang1986}, and phototaxis in mutated Halobacteria~\cite{Hoff1997}. In \textit{S. typhimurium} chemotaxis, this alternative response can be caused by one of a number of single point mutations~\cite{Khan1978}. Interestingly, this behavior was also found in wild-type, aerobically-grown \textit{R. sphaeroides}~\cite{Packer2000}. This response seems paradoxical, as the bacterium might be expected to run in the direction of decreasing attractant concentration and tumble more when it detects an increase in attractant concentration, leading to the accumulation of bacteria away from the attractant. In addressing these puzzling results, Goldstein and Soyer demonstrated the chemotactic efficacy of a non-adaptive ``inverted'' response~\cite{Goldstein2008, Soyer}. With this strategy, bacteria respond to the absolute attractant concentration, tumbling more and therefore maintaining their position in regions of higher attractant concentration. Although less effective than the adaptive response, the inverted response requires only low receptor sensitivity, and could function in the absence of effective receptors by coupling to the cell metabolism~\cite{Egbert2010}. As we show, there are discrepancies between the inverted response and the response observed in aerobically-grown \textit{R. sphaeroides}.

Why do different bacteria exhibit different chemotactic responses? One possible reason is that different bacteria have evolved for different environments. For example, while \textit{E. coli} might be expected to inhabit a resource-rich environment, marine bacteria experience a harsh environment in which attractant is localized in short-lasting patches~\cite{Blackburn1998} with attractant concentration inside patches being 3 to 6 orders of magnitude higher than outside~\cite{Stocker2012}. This has led to a number of marine-specific evolutionary adaptations such as higher running speed in \textit{Pseudomonas haloplanktis}~\cite{Stocker2008} and run-and-reverse (as opposed to run-and-tumble) chemotaxis in over 70\% of marine bacterial species~\cite{Johansen2002}. Unfortunately, most experimental and theoretical studies to date consider chemotaxis in response to step functions or simple gradients~\cite{Ahmed2010}, providing limited insights to how chemotaxis would function in different types of environments. In particular, there have few studies~\cite{Blackburn1999, Celani2010} analyzing chemotactic strategies in stochastic environments, which are likely to be the most important environment during the evolution of a chemotactic response.

In this work we construct a model of a stochastic attractant distribution. By adjusting the manner in which attractant concentrations vary in time and space, the attractant distribution can mimic a range of environments that one might expect to find in nature. We use this model to study how performance and optimal properties of various chemotactic strategies vary as a function of environmental conditions. In particular, we describe a new type of chemotactic strategy called the speculator response. It differs from the adaptive response in that the tumbling rate increases with increasing attractant concentration; furthermore, the bacterium makes temporal comparisons of attractant concentration which distinguishes the strategy from the inverted response. We demonstrate the effectiveness of the speculator strategy and its remarkable match to the paradoxical response seen in wild-type, aerobically-grown \textit{R. sphaeroides}~\cite{Packer2000}.

\section*{Results}

\begin{figure*}[htb]
{\rotatebox{0}{{\includegraphics[width=\textwidth]{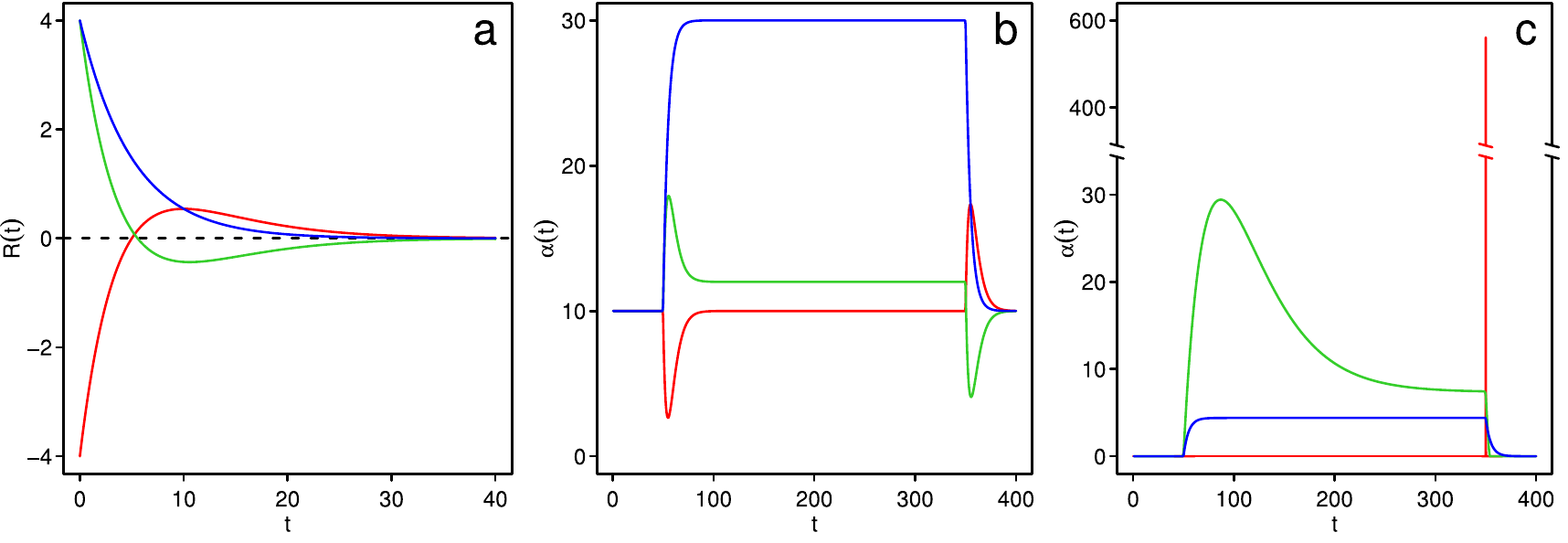}}}}
    \caption{{\bf Illustrative examples of the response function and the corresponding timecourses of $\alpha$ in response to attractant step functions. Timecourses of $\alpha$ in response to attractant step functions for optimized responses.} a) Illustrative examples of the response function for adaptive (red), inverted (blue) and speculator (green) response. The parameters used are $A = -20$, $B = 20$ and $\tau = 5$ for adaptive, $A = 20$, $B = 0$ and $\tau = 5$ for inverted, and $A = 20$, $B = -18$ and $\tau = 5$ for speculator response. b) Changes in $\alpha$ in response to step changes in attractant concentration for the responses from part a). $\alpha_0 = 10$ for all responses. Attractant (concentration of 1) is added at $t = 50$ and removed at $t = 350$. c) Changes in $\alpha$ in response to step changes in attractant concentration for adaptive (red), inverted (blue) and speculator (green) responses optimized for $T = 10^4$ and $L = 100$ ($\alpha_{0} = 0.0084$, $A = -1500$, $B = 1500$ and $\tau = 0.020$ in the adaptive response, $\alpha_{0} = 0.0063$, $A = 4.4$, $B = 0$ and $\tau = 5.0$ in the inverted response, and $\alpha_{0} = 0.0089$, $A = 74$, $B = -67$ and $\tau = 33$ in the speculator response). Attractant (concentration of 1) is added at $t = 50$ and removed at $t = 350$. Note the change of scale on the $y$-axis.}
    \label{fig:alpha}
\end{figure*}

To model chemotaxis, we consider a single bacterium in a one-dimensional space with periodic boundary conditions and a distribution of attractant. The bacterium can run to the left or right, or tumble. $\alpha$ and $\beta$ denote the rates at which the bacterium starts and stops tumbling. While $\beta$ is assumed to be constant, the basal rate $\alpha_{0}$ is modulated by the chemotactic response of the bacterium to the experienced attractant concentrations. In particular, at time $t$,
\begin{equation}
    \alpha(t) = \max \left( 0, \, \alpha_{0} + \int_{-\infty}^{t} R(t - t') c(x_B(t'), t') \, \mathrm{d} t' \right) \label{eq:alpha}
\end{equation}
where $R(t)$ is the chemotactic response function and $c(x_B(t), t)$ is the attractant concentration that the bacterium experiences at position $x_B(t)$ at time $t$~\cite{Clark2005}. In contrast to~\cite{Clark2005}, we do not assume deviations from $\alpha_0$ to be small. $R(t)$ is represented as $(A/\tau + Bt/\tau^2) \exp(-t/\tau)$ where $\tau$ controls the memory length, i.e.\ how far back in the past the bacterium ``remembers'' attractant concentrations, and $A$ and $B$ together determine the sensitivity and the characteristics of the response: adaptive, inverted, or speculator. In the adaptive response, $A$ and $B$ are constrained such that $A < 0$ and $B = -A$. This gives rise to a response function shown in red in Fig.~\ref{fig:alpha}a that has a positive and a negative lobe. The red curve in Fig.~\ref{fig:alpha}b illustrates the changes in $\alpha$ due to attractant addition and removal when a response function of this type is used. When $c(x_B(t), t)$ is increasing in time, such as when attractant is added, the negative lobe of $R(t - t') c(x_B(t'), t')$ has a larger area than the positive lobe, making the integral in Eq.~(\ref{eq:alpha}) negative, leading to an $\alpha$ that is smaller than $\alpha_{0}$, resulting in a decrease in tumbling, as the red curve in Fig.~\ref{fig:alpha}b shows at $t = 50$. The opposite happens when attractant is removed, resulting in an increase in tumbling at $t = 350$. The constraint $B = -A$ results in equal areas of the positive and negative lobes of $R(t)$, ensuring perfect adaptation and a basal tumbling rate ($\alpha = \alpha_0$) for $50 < t < 350$. In the inverted response (Figure~\ref{fig:alpha}a, blue curve), $A > 0$ and $B = 0$, leading to a single-lobe response function. This results in a higher tumbling rate in the presence of attractant and a lower rate when attractant is absent, as the blue curve in Fig.~\ref{fig:alpha}b shows. We also investigate a new type of response which we name the ``speculator'' response for reasons explained below. In the speculator response, $A > 0$ and $B < 0$, leading to a double-lobe response function (Figure~\ref{fig:alpha}a, green curve) that looks roughly like the negative of the adaptive response function (Figure~\ref{fig:alpha}a, red curve). This causes increased (decreased) tumbling when $c(x_B(t), t)$ is increasing (decreasing) in time (Figure~\ref{fig:alpha}b, green curve). The constraint of perfect adaptation is relaxed in the speculator response, so the areas of the positive and negative lobes are unequal (Figure~\ref{fig:alpha}a, green curve). This allows the steady-state $\alpha$ in the presence of constant attractant concentration to be different from $\alpha_{0}$, as shown for times $100 < t < 350$ (Figure~\ref{fig:alpha}b, green curve). The double-lobe response functions of adaptive and speculator response cause bacteria to make temporal comparisons of attractant concentration, while the lack of perfect adaptation in the inverted and speculator responses causes bacteria to respond to absolute attractant concentrations. Note that Fig.~\ref{fig:alpha}a and Fig.~\ref{fig:alpha}b are purely illustrative; they do not reflect real or optimized responses.

\begin{figure}[h]
{\rotatebox{0}{{\includegraphics[width=0.5\textwidth]{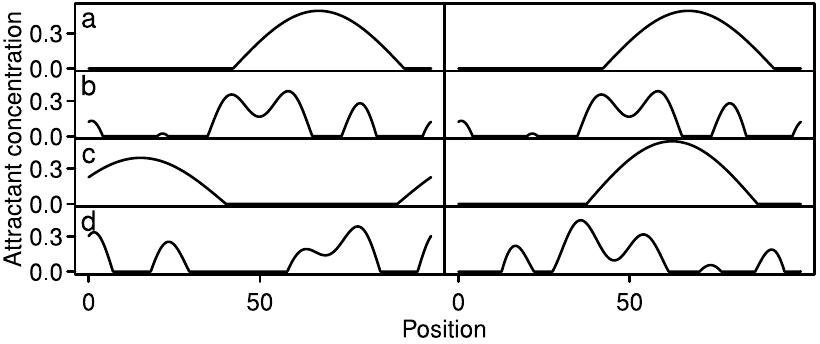}}}}
    \caption{{\bf Examples of the stochastic attractant distribution at different combinations of $T$ and $L$.} Every row corresponds to a different combination. The left (right) panels show the distribution at time $t = 0$ ($t = 100$) in the simulation. a) $T = 10^4$, $L = 100$. b) $T = 10^4$, $L = 20$. c) $T = 100$, $L = 100$. d) $T = 100$, $L = 20$.}
    \label{fig:food}
\end{figure}

The one-dimensional virtual world in which the bacterium moves contains a stochastic attractant distribution which varies in both time and space. Two parameters, $T$ (correlation time) and $L$ (correlation length), determine the dynamics of the attractant distribution. $T$ is the timescale on which attractant concentrations change, while $L$ determines the distance between peaks of attractant concentration; the shorter $L$, the more numerous and narrow the peaks are and the shorter the distances between them. The average amount of attractant available in the world is independent of $T$ and $L$. Fig.~\ref{fig:food} shows how the distribution looks at different combinations of $T$ and $L$. \nameref{S1_Video} illustrates the distribution dynamics as a function of $T$ and $L$.

The framework described above allows us to assess the performance of a chemotactic response characterized by the response parameters $\alpha_{0}$, $\beta$, $A$, $B$ and $\tau$ at a chosen combination of attractant distribution parameters $T$ and $L$. Performance, or fitness, of a response, is equal to the average cell division rate, which we approximate as the inverse of the time it takes the bacterium to experience a specified amount of attractant. For any chemotactic strategy in any stochastic environment, we can optimize the response parameters to maximize the bacterial fitness. Performing this optimization for different strategies (by applying appropriate constraints on $A$ and $B$) under different combinations of $T$ and $L$ allows us to explore the performance of the different strategies and how this performance varies with $T$ and $L$.

\begin{figure}[h]
{\rotatebox{0}{{\includegraphics[width=0.5\textwidth]{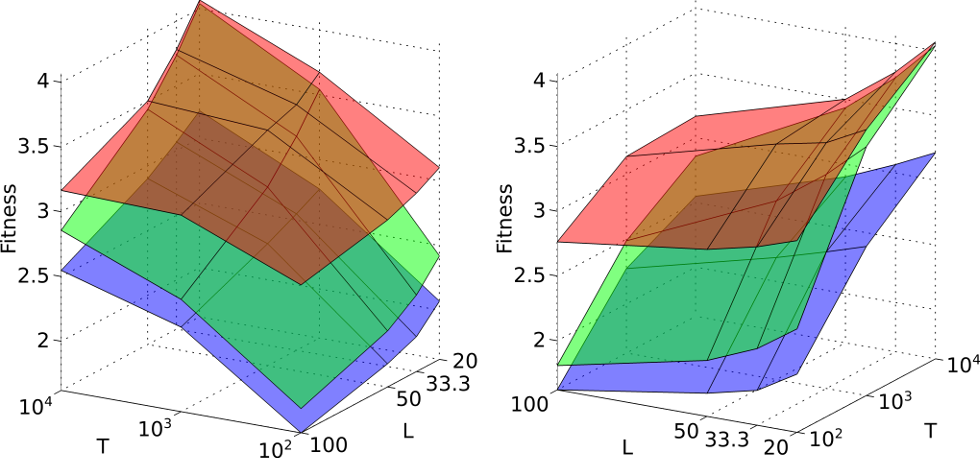}}}}
    \caption{{\bf Fitness of optimized adaptive (red), speculator (green) and inverted (blue) response as a function of $T$ and $L$.} The two panels show different views of the same plot. For each chemotactic strategy and combination of $T$ and $L$, fitnesses are averaged over the last 600 generations of up to 3 replicate simulations.}
    \label{fig:fitness}
\end{figure}

Figure~\ref{fig:fitness} shows the optimal fitnesses of adaptive, inverted and speculator responses as a function of $T$ and $L$. Fitnesses are scaled by the fitness of a non-chemotaxing bacterium, whose fitness is independent of $T$ and $L$: a bacterium with a relative fitness of 4 therefore takes 4 times less time to experience the same amount of attractant than a non-chemotaxing bacterium. In all strategies, fitness increases with increasing $T$ and decreasing $L$. As $T$ increases, attractant concentrations change more slowly, making it easier for bacteria to track attractant peaks. At shorter $L$, fitness is higher because there are more attractant peaks and they are closer to one another. This means that if a bacterium loses track of a peak, or a peak diminishes in amplitude over time, the bacterium only needs to travel a short distance to reach another peak.

In addition to the previously studied adaptive and inverted responses, we characterize a novel chemotactic strategy. This ``speculator'' response, despite its seemingly paradoxical nature, is more fit under all studied conditions than the inverted response, although less fit than the adaptive response. Interestingly, at $T = 10^4$, $L = 20$, the fitness of the speculator response is nearly identical to the fitness of the adaptive response. To understand the mechanism of the speculator response, we consider the optimal values of response parameters (Table~\ref{tab:response_parameters}). The lack of perfect adaptation (optimal $|A| > |B|$) means that the bacterium will more often start to tumble when the attractant concentration is high, as shown in green in Fig.~\ref{fig:alpha}c; the low value of $\beta$ means that the bacterium will then continue tumbling, remaining in the region of high attractant. Consequently, the speculator response, like the inverted response (Figure~\ref{fig:alpha}c, blue curve), results in frequent long tumbles at high attractant concentrations. In contrast to the inverted response, the double-lobe response function of the speculator response results in a tumbling rate sensitive to the rate of change of the attractant. The long memory of the speculator response (large $\tau$) allows sensitivity to long-term trends; this sensitivity, combined with the double-lobe response function, results in two important dynamical properties. Firstly, the bacterium compares recent attractant concentrations with a long-term average, tumbling more when the recent past is more favorable than the average, and therefore maintaining its position in regions of higher attractant concentration. Secondly, the bacterium is able to sense improving and worsening conditions at its current location. In particular, a decline in the attractant concentration results in a decrease in $\alpha$, allowing the bacterium to swim away from a peak when conditions are changing for the worse. Swimming away leads to a further decrease in $\alpha$, setting a feedback loop in motion, resulting in continued swimming until a new optimum is reached. The speculator response is therefore analogous to the behavior of investors in financial markets: when the current performance is lower than the average, or when investment values are falling, speculators seek higher returns by abandoning their current position and investing elsewhere---hence the name ``speculator'' response. The behavior of the speculator response, compared with the adaptive and inverted responses, is illustrated in~\nameref{S2_Video}.


\begin{table*}[!ht]
        \caption{{\bf Range of optimal response parameters at $T = 10^3$, $L = 50$.} 3 replicate simulations are run for each chemotactic strategy at $T = 10^3$, $L = 50$. For a given strategy and response parameter, mean values of the parameter are calculated separately in each replicate simulation by averaging the values of the parameter over the last 600 generations. Each range in the table is composed of the lowest and highest mean values obtained.}
        \begin{tabular}{|l|c|c|c|c|c|}
            \hline
             & $\tau$ & $\alpha_{0}$ & $\beta$ & $A$ & $B/A$ \\
            \hline
            Adaptive & (0.072, 0.27) & (0.012, 0.062) & (10, 47) & ($-11000$, $-5700$) & $-1$ \\
            \hline
            Inverted & (5.6, 8.3) & (0.0024, 0.0049) & (0.098, 0.68) & (2.0, 6.2) & 0 \\
            \hline
            Speculator & (39, 45) & (0.0030, 0.046) & (0.086, 0.44) & (47, 96) & ($-0.86$, $-0.82$) \\
            \hline
        \end{tabular}
            
        \label{tab:response_parameters}
\end{table*}

\begin{figure}[h]
{\rotatebox{0}{{\includegraphics[width=0.5\textwidth]{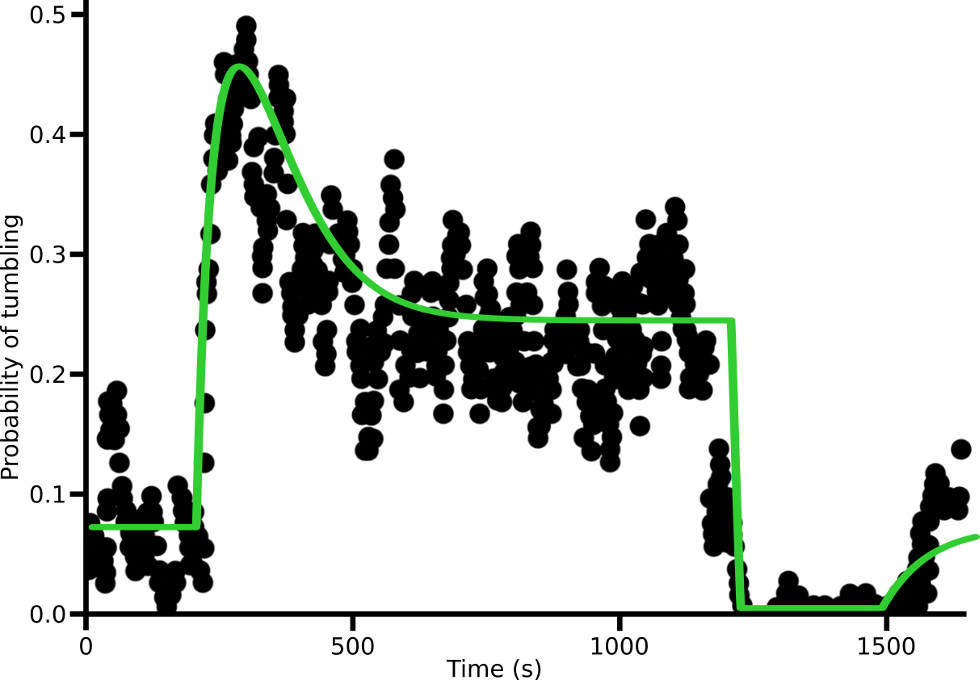}}}}
    \caption{{\bf Curve-fit of our model of the speculator response (green curve) to the digitized data of~\cite{Packer2000} (black circles).} The response is described by the following response parameters: $\alpha_{0} = 0.074 \, \mathrm{s}^{-1}$, $\beta = 0.034 \, \mathrm{s}^{-1}$, $A = 1300 \, \mathrm{mM}^{-1}\mathrm{s}^{-1}$, $B = -1000 \, \mathrm{mM}^{-1}\mathrm{s}^{-1}$ and $\tau = 71 \, \mathrm{s}$. Attractant concentration is set to 0.001~mM in correspondence to~\cite{Packer2000}. The curve-fit is obtained by optimizing the response parameters of the speculator response to minimize the least-squares fit between the model and the digitized data. The data are digitized in MATLAB using the function imfindcircles~\cite{Matlab}.}
    \label{fig:curve_fit}
\end{figure}

Significantly, the time course of $\alpha$ in the speculator response closely matches the time course of probability of tumbling in aerobically-grown \textit{R. sphaeroides} (see Fig.~2A in~\cite{Packer2000}). Fig.~\ref{fig:curve_fit} shows a curve-fit of our model of the speculator response to the digitized data of~\cite{Packer2000}. The closeness of the fit provides strong evidence that aerobically-grown \textit{R. sphaeroides} uses the speculator response to respond to Na--succinate. Experimental results show that aerobically-grown \textit{R. sphaeroides} performs well in swarm plates~\cite{Packer2000}, demonstrating the efficacy of this response.

\begin{figure}[h]
{\rotatebox{0}{{\includegraphics[width=0.5\textwidth]{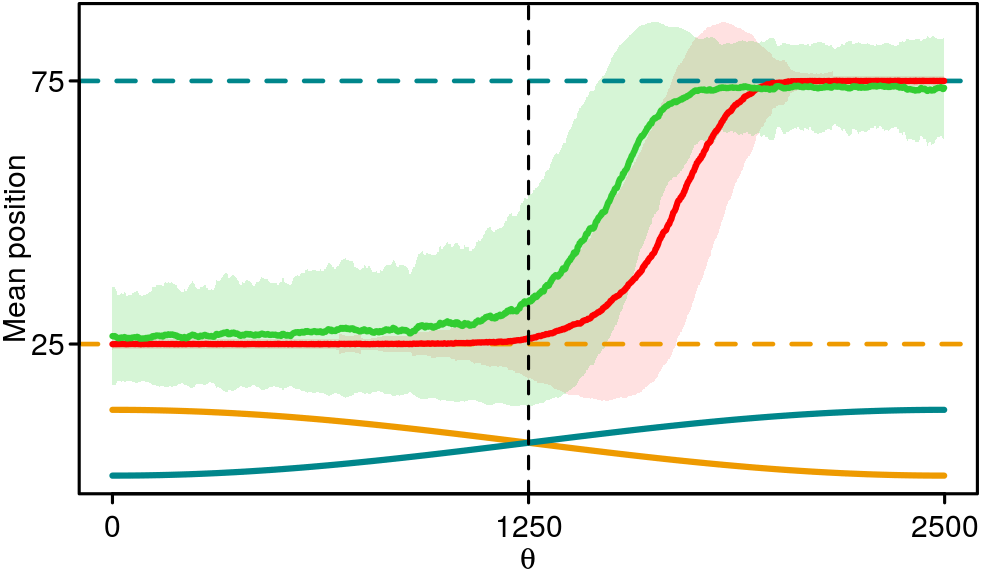}}}}
    \caption{{\bf Mean position of bacteria performing adaptive (red curve) or speculator (green curve) response as a function of $\theta$, the phase of the oscillations.} Shading shows the standard deviation of the position. The Gaussians are centered at positions 25 (dashed gold line) and 75 (dashed turquoise line) and have a standard deviation of 3. Amplitudes of the Gaussians (not to scale, maximum amplitude is 1) are shown as a function of $\theta$ in the bottom part of the figure for the Gaussian at position 25 (gold curve) and 75 (turquoise curve). The parameters used are $\alpha_0 = 0.0084$, $\beta = 54$, $A = -1526$ and $\tau = 0.020$ for the adaptive and $\alpha_0 = 0.0089$, $\beta = 0.056$, $A = 74$, $B = -67$ and $\tau = 33$ for the speculator response.}
    \label{fig:exp_vs_exp}
\end{figure}

As Fig.~\ref{fig:fitness} shows, at $T = 10^4$, $L = 20$, fitnesses of the adaptive and speculator responses are very similar despite the different mechanisms behind their chemotactic strategies. To better understand these strategies, we create a simple attractant distribution which consists of two Gaussians (at positions 25 and 75 in a world with a length of 100) oscillating in amplitude out of phase with each other: when one Gaussian is at full amplitude, the other has amplitude of zero. Amplitude, period of oscillation and width of the Gaussians are roughly matched to $T = 10^4$, $L = 20$ of the stochastic attractant distribution. For each of the two chemotactic strategies, we take a bacterium optimized for $T = 10^4$, $L = 20$ of the stochastic attractant distribution and simulate its movement in the virtual world with the two Gaussians. Fig.~\ref{fig:exp_vs_exp} shows the mean position of the bacteria as a function of $\theta$, the phase of the oscillations. Between $\theta = 0$ and $\theta = 1250$ (the period is 5000) the left Gaussian at position 25 is higher than the right Gaussian at position 75, but is decreasing. In the adaptive response (red curve), the bacterium is close to the top of this Gaussian during this period. The bacterium shows little movement toward the right Gaussian at position 75 until the right Gaussian is significantly higher i.e.\ $\theta > 1250$. In the speculator response (green curve), the bacterium cannot track the top of the Gaussian as well as in the adaptive response, as evidenced by the large standard deviation around position 25 (green shading). However, the bacterium more quickly adapts to the changing attractant levels, leaving the declining left Gaussian and moving towards the growing right Gaussian sooner.

The strengths of the adaptive and speculator responses therefore lie in exploitation and exploration, respectively. In the adaptive response, the bacterium can track the top of a peak efficiently while in the speculator response, the bacterium is better at leaving the declining peak and finding the increasing peak. The exploitation behavior of the adaptive response is analogous to a hill-climbing algorithm, which efficiently finds, but may get stuck at, a local optimum. The exploration behavior of the speculator response is more analogous to a Monte Carlo search algorithm in that the bacterium may leave a peak in search of a higher peak at the cost of its ability to track the peak top efficiently. This explains the trend in Fig.~\ref{fig:fitness}: for large $L$, the number of attractant peaks is small, and exploiting a given peak is more important than exploring new peaks. Under these conditions, the adaptive response is significantly more effective than the speculator response. At short $L$, there are multiple peaks in the environment, each of which has a different amplitude. Under such conditions, the exploration behavior of the speculator response allows the bacterium to locate higher peaks, while the exploitation behavior of the adaptive response may lead to the bacterium tracking a suboptimal peak. At $T = 10^4$, $L = 20$, the two strategies are approximately equally effective, giving rise to very similar fitnesses (Figure~\ref{fig:fitness}).

We next consider the optimal values of the response parameters. In the adaptive response, $\tau$ (the memory length) is very short (Table~\ref{tab:response_parameters}), allowing the bacterium to quickly adjust to small displacements from attractant optima. $\beta$, the rate at which the bacterium stops tumbling, is quite large, corresponding to short-lasting tumbles characteristic of chemotaxis in \textit{E. coli}~\cite{Berg1972}. High sensitivity (large $|A|$ and $|B|$) is necessary for the bacterium to respond to small differences in attractant concentration characteristic of small displacements from the top of an attractant peak. High sensitivity is responsible for the high $\alpha$ when the attractant is removed at $t = 350$ in Fig.~\ref{fig:alpha}c (red curve). Optimal $\alpha_0$, the tumbling rate in the absence of attractant (or under constant attractant in case of perfect adaptation), is very low in all strategies, as is evident from Fig.~\ref{fig:alpha}c. Low $\alpha_0$ enables bacteria to run persistently in order to find regions with more favorable conditions more quickly. The near-zero value of $\alpha_0$ removes the possibility of $\alpha$ going below $\alpha_0$, eliminating the response to increasing attractant in the adaptive response (Figure~\ref{fig:alpha}c, red curve, $t = 50$).

In the inverted response, the bacterium tumbles more at higher concentrations of attractant (Figure~\ref{fig:alpha}c, blue curve). $\tau$ is longer than for the adaptive response, allowing the bacterium to integrate over short-term fluctuations. In both inverted and speculator responses, $\beta$ is much lower than in the adaptive response, resulting in significantly longer tumbles. This is central to the strategies, as it is the persistence of position when tumbling that allows bacteria to stay in regions of high attractant concentration. The sensitivity is lower than in the adaptive response, in agreement with simple models showing that the inverted response is optimized by lower sensitivity~\cite{Soyer}. Sensitivity needs to be tailored to the range of attractant concentrations the bacterium experiences: if it is too low, the bacterium will run past high concentrations of attractant; if it is too high, the bacterium will tumble at low concentrations of attractant, never reaching higher concentrations.

\section*{Discussion}
In this work we describe a new chemotactic strategy, termed the speculator response, in which the bacterium compares the current attractant concentration with a long-term average; if the current concentration is higher than the long-term average, the bacterium tumbles persistently to maintain its position. On the other hand, declines in the current concentration will increase the probability that the bacterium will swim away to a higher peak. By considering stochastic attractant distributions, we show that under slowly-changing but spatially complex attractant concentrations (large $T$, small $L$), the speculator response is almost as efficient at co-localizing with attractant as the adaptive response of \textit{E. coli} (Figure~\ref{fig:fitness}). While the adaptive response achieves high fitness by accurately tracking the top of an attractant peak, the speculator response enables the bacterium to explore its environment and find higher peaks more efficiently (Figure~\ref{fig:exp_vs_exp}).

The speculator response closely matches the response observed in wild-type, aerobically-grown \textit{R. sphaeroides} (Figure~\ref{fig:curve_fit}). The optimized response parameters from our simulations are in arbitrary units, and cannot be directly compared with those obtained by the fit to the wild-type response (Figure~\ref{fig:curve_fit}). Interestingly, however, the ratio of $B$ to $A$ which quantifies the extent of departure from perfect adaptation ($B / A = -1$ corresponds to perfect adaptation) is similar between the optimized values obtained from the simulations and the response observed in aerobically-grown \textit{R. sphaeroides} ($-0.86$ and $-0.82$, respectively). Furthermore, we can acquire a rough estimate of the ratio of $\tau_{S} / \tau_{A}$ (where $\tau_{S}$ and $\tau_{A}$ are the values of $\tau$ in the speculator and adaptive responses) by comparing the values of $\tau$ in aerobically-grown \textit{R. sphaeroides} (Figure~\ref{fig:curve_fit}) and wild-type \textit{E. coli}~\cite{Segall1986}. This ratio ($71 / 1 = 71$) is of similar order of magnitude to the ratio for the optimized simulated responses ($43 / 0.20 = 215$), despite the multitude of differences between wild-type \textit{R. sphaeroides} and \textit{E. coli}.

The optimized adaptive response possesses high sensitivity (large $|A|$ and $|B|$; Table~\ref{tab:response_parameters}) consistent with experimental results from \textit{E. coli}~\cite{Sourjik2002}. Furthermore, $\beta$, the rate at which the bacterium stops tumbling, is high, which is in line with the short tumbles observed in real bacteria~\cite{Berg1972}. In contrast to real bacteria, the optimized bacteria have a lower $\alpha_0$, and thus tumble less than real bacteria when attractant concentration is increasing (Figure~\ref{fig:alpha}c, red curve, $t = 50$). This may be an artifact of modeling chemotaxis in a one-dimensional environment: in a three-dimensional environment, tumbling may assist the bacterium in finding even steeper paths to attractant optima.

Our model does not take into account the motility-associated energy costs of the different chemotactic strategies. For instance, \textit{R. sphaeroides} does not actively tumble, but rather stops running and lets rotational diffusion generate the re-orientation, reducing the costs of strategies that involve longer tumbles~\cite{Mitchell2002}. The speculator response therefore might have emerged partly because \textit{R. sphaeroides} uses rotational diffusion to achieve tumbling. Alternatively, rotational diffusion might have emerged in response to the bacterium using a strategy that involves long tumbles. In addition, by necessity we are confined to a relatively small range of $T$ and $L$; other conditions might exist (such as larger $T$ and smaller $L$) that would favor the speculator response even more.

Our approach differs from that of other studies in that we consider realistic attractant distributions and extended tumbling times. The latter is essential for the speculator response to work as it allows bacteria to maintain their position in regions of high attractant concentration. Previous studies~\cite{Schnitzer1993, Clark2005, DeGennes2004} modeled tumbles as instantaneous after chemotaxis in \textit{E. coli}~\cite{Berg1972}, however, experimental evidence from other bacterial species shows longer tumbling times~\cite{Gotz1987, Armitage1987}. Our results add to the growing body of evidence that extended tumbles allow for emergence of other modes of chemotaxis~\cite{Goldstein2008, Soyer, Kafri2008}.

Most studies to date considered chemotaxis in response to step functions or simple gradients~\cite{Ahmed2010}. While this is important for our understanding of the basic mechanisms of chemotaxis, we should recognize that chemotactic strategies were inevitably shaped by the environments the bacteria inhabited. For example, studies in marine bacteria unearthed specific adaptations to marine environments~\cite{Blackburn1999, Stocker2008, Johansen2002}, highlighting the need to study chemotaxis in the context of realistic attractant distributions. Here, we propose a model of a stochastic attractant distribution which allows us to compare the performance of various chemotactic strategies under different environments and study how optimal properties of chemotactic responses change as a function of environmental conditions. This can also help us characterize the environmental conditions based on the strategies that have evolved. Further characterization of natural environments~\cite{Blackburn1998} will allow theorists to construct more detailed attractant distributions and advances in microfluidics technologies will enable these environments to be reconstructed in laboratory settings~\cite{Ahmed2010}.

\section*{Methods}
\subsection*{Stochastic attractant distribution}
We generate our stochastic attractant distribution by summing over cosine and sine modes with different mode numbers $p$ so that the concentration at position $x$ and time $t$ along the virtual world is calculated as
\begin{equation}\label{Eq:Food}
c(x, t) = \max \left( 0, \; \sum_{p = 1}^{p^*} X_p (t) \cos \xi_p + Y_p (t) \sin \xi_p \right)
\end{equation}
where $X_p (t)$ and $Y_p (t)$ are stochastic weights, $l$ is the length of the one-dimensional virtual world ($l = 100$), $\xi_p = 2 \pi p x / l$ and $p^*=l/L$ is the largest mode included in the sum above. $X_p (0) = Y_p (0) = 0$ for all $p$s, and are updated at intervals of $\Delta t_c = T / 100$ according to:
\begin{equation}\label{Eq:ModeAmplitudes}
X_p (t + \Delta t_c) = X_p (t) (1 - \Delta t_c / T) + \eta_p(t) \sqrt{\frac{2 \Delta t_c}{Tp^*}}
\end{equation}
where $\eta_p(t)$ is a white noise Gaussian random process ($\langle \eta_p(t)\eta_q(t')\rangle=\delta(t-t')\delta_{pq}$), generated by a random number sampled from a normal distribution with mean 0 and variance 1. A similar expression is used for $Y_p (t + \Delta t_c)$. By construction, this results in a Markov process with correlation time $T$ and approximate correlation length $L = l / p^*$.

\subsection*{Chemotaxis}
The attractant distribution is equilibrated for a period of at least $T$. Before a bacterium is introduced, $\alpha$ is initialized based on the equilibrated attractant distribution. The bacterium is then released and the state of the bacterium (whether it is running or tumbling) is updated every $\Delta t_B = \min (T, L/v, \tau / 20)$ where $v$ is the speed of the bacterium when running ($v = 1$). A Monte Carlo scheme is used to decide whether the bacterium starts tumbling (running) given that it was running (tumbling) previously, assuming first-order dynamics of a 2-state system. When the bacterium stops tumbling, it starts running left or right with equal probability.

\subsection*{Optimization}
Mutagenesis followed by selection constitute one generation of the optimization. In the first generation, all response parameters are initialized randomly from a uniform distribution between 0 and 1 (but see below). In the adaptive response, only $B$ is initialized and mutated, $A$ is set to $-B$ (at $T = 10^4$, $B$ is initialized between 1 and 10). In the speculator response, $B$ is initialized randomly between 0 and $-1$. In every generation, one response parameter is chosen at random and mutated. Parameters are mutated on a log scale by a transformation $\exp(\log_e(a) + r)$ where $a$ is the parameter being mutated and $r$ is a random number sampled from a uniform distribution between $-0.2$ and 0.2. Further constraints on response parameter values are imposed for reasons of computational tractability: $\alpha_0 > 10^{-3}$ in adaptive and inverted response, $A > \exp(-1)$ and $|B| > \exp(-1)$ in speculator response, $\tau > 0.01$ in adaptive response.

After mutagenesis, the fitnesses of responses described by the wild-type and mutant response parameters are determined. This is achieved by letting 10 identical wild-type and 10 identical mutant bacteria explore the virtual world with the stochastic attractant distribution. Each of the 10 wild-type bacteria is subjected to an attractant distribution initialized with a different random seed; the attractant distributions are then re-used for the 10 mutants. (As the attractant distribution is stochastic, estimates of response fitness are stochastic too. This scheme of competing the wild-type and mutant with the same attractant distributions is thus used to ensure that lucky mutants do not fix.) Each of the bacteria is run until it experiences $50T$ attractant units. $D_{\mathrm{w}, i}$ ($D_{\mathrm{m}, i}$) denotes the time it took the $i$-th wild-type (mutant) to experience the specified amount of attractant. Fitness of response $k = \{\mathrm{w}, \mathrm{m}\}$, $F_{k}$, is then calculated as $T \langle \frac{1}{D_{k, i}} \rangle_i$ averaged over the bacteria.

Once fitnesses are determined, the probability of acceptance of the mutation, $p_{\mathrm{m}}$, is calculated using the Metropolis-Hastings algorithm: $p_{\mathrm{m}} = 1$ if $F_{\mathrm{m}} \geq F_{\mathrm{w}}$ and $p_{\mathrm{m}} = \exp( (F_{\mathrm{m}} - F_{\mathrm{w}}) / (U F_{\mathrm{w}}) )$ otherwise. $U$, the temperature, is constant at 0.005. Simulations are run until the fitness stops increasing and stays constant for at least 600 generations. 3 replicate simulations are run for each chemotactic strategy and combination of $T$ and $L$.

\section*{Supporting Information}

\subsection*{S1 Video}
\label{S1_Video}
{\bf Dynamics of the attractant distribution for a) $T = 10000$, $L = 100$, b) $T = 10000$, $L = 20$, c) $T = 100$, $L = 100$, a) $T = 100$, $L = 20$.}

\subsection*{S2 Video}
\label{S2_Video}
{\bf Dynamics of optimized adaptive (a), inverted (b) and speculator (c) strategies under $T = 10000$ and $L = 20$.} The parameters used are: $\alpha_0 = 0.0065$, $\beta = 4.2$, $A = -2100$, $B = 2100$, $\tau = 0.016$ in adaptive, $\alpha_0 = 0.0016$, $\beta = 0.048$, $A = 3.7$, $B = 0$, $\tau = 5.8$ in inverted and $\alpha_0 = 0.0089$, $\beta = 0.056$, $A = 74$, $B = -67$, $\tau = 33$ in speculator response. In the adaptive response, the bacterium swims up attractant gradients and tumbles when it experiences a decrease in attractant concentration. This leads to an oscillatory behavior around peak maxima. In the inverted response, tumbling rate increases with increasing attractant concentration. Response sensitivity is optimized such that the bacterium tumbles most persistently at attractant concentrations which correspond to typical values at attractant maxima. However, this means that the bacterium can get stuck at sub-optimal concentrations on large peaks. The speculator response compares the current concentration of attractant with a long term average. If the current concentration is greater than this average, the bacterium tumbles more. If the current concentration is lower than the average, or declining, the bacterium swims away, leaving the peak to search for higher attractant concentrations. The bacterium will typically run past peaks if their amplitude is lower than the peak it just left.

\section*{Acknowledgments}
We would like to thank Orkun Soyer and Christopher Monit for helpful discussions and funding from the Medical Research Council, U.K (funding reference U117573805).


\begin{thebibliography}{10}
\providecommand{\url}[1]{\texttt{#1}}
\providecommand{\urlprefix}{URL }
\expandafter\ifx\csname urlstyle\endcsname\relax
  \providecommand{\doi}[1]{doi:\discretionary{}{}{}#1}\else
  \providecommand{\doi}{doi:\discretionary{}{}{}\begingroup
  \urlstyle{rm}\Url}\fi
\providecommand{\bibAnnoteFile}[1]{%
  \IfFileExists{#1}{\begin{quotation}\noindent\textsc{Key:} #1\\
  \textsc{Annotation:}\ \input{#1}\end{quotation}}{}}
\providecommand{\bibAnnote}[2]{%
  \begin{quotation}\noindent\textsc{Key:} #1\\
  \textsc{Annotation:}\ #2\end{quotation}}
\providecommand{\eprint}[2][]{\url{#2}}

\bibitem{Wuichet2010}
Wuichet K, Zhulin IB (2010) {Origins and diversification of a complex signal
  transduction system in prokaryotes}.
\newblock Science Signaling 3: ra50.
\input{Wuichet2010}

\bibitem{Hamer2010}
Hamer R, Chen PY, Armitage JP, Reinert G, Deane CM (2010) {Deciphering
  chemotaxis pathways using cross species comparisons}.
\newblock BMC Systems Biology 4: 3.
\input{Hamer2010}

\bibitem{Poole1988}
Poole PS, Armitage JP (1988) {Motility response of Rhodobacter sphaeroides to
  chemotactic stimulation}.
\newblock Journal of Bacteriology 170: 5673--9.
\input{Poole1988}

\bibitem{Armitage1997}
Armitage JP (1997) {Behavioural responses of bacteria to light and oxygen}.
\newblock Archives of Microbiology 168: 249--61.
\input{Armitage1997}

\bibitem{Dang1986}
Dang CV, Niwano M, Ryu JI, Taylor BL (1986) {Inversion of Aerotactic Response
  in Escherichia coli Deficient in cheB Protein Methylesterase}.
\newblock Journal of Bacteriology 166: 275--80.
\input{Dang1986}

\bibitem{Bespalov1996}
Bespalov VA, Zhulin IB, Taylor BL (1996) {Behavioral responses of Escherichia
  coli to changes in redox potential}.
\newblock Proceedings of the National Academy of Sciences of the United States
  of America 93: 10084--9.
\input{Bespalov1996}

\bibitem{Hoff1997}
Hoff WD, Jung KH, Spudich JL (1997) {Molecular mechanism of photosignaling by
  archaeal sensory rhodopsins}.
\newblock Annual Review of Biophysics and Biomolecular Structure 26: 223--58.
\input{Hoff1997}

\bibitem{Khan1978}
Khan S, Macnab RM, DeFranco AL, {Koshland Jr} DE (1978) {Inversion of a
  behavioral response in bacterial chemotaxis: explanation at the molecular
  level}.
\newblock Proceedings of the National Academy of Sciences of the United States
  of America 75: 4150--4.
\input{Khan1978}

\bibitem{Packer2000}
Packer HL, Armitage JP (2000) {Inverted behavioural responses in wild-type
  Rhodobacter sphaeroides to temporal stimuli}.
\newblock FEMS Microbiology Letters 189: 299--304.
\input{Packer2000}

\bibitem{Goldstein2008}
Goldstein RA, Soyer OS (2008) {Evolution of taxis responses in virtual
  bacteria: non-adaptive dynamics}.
\newblock PLoS Computational Biology 4: e1000084.
\input{Goldstein2008}

\bibitem{Soyer}
Soyer OS, Goldstein RA (2011) {Evolution of response dynamics underlying
  bacterial chemotaxis}.
\newblock BMC Evolutionary Biology 11: 240.
\input{Soyer}

\bibitem{Egbert2010}
Egbert MD, Barandiaran XE, {Di Paolo} EA (2010) {A minimal model of
  metabolism-based chemotaxis}.
\newblock PLoS Computational Biology 6: e1001004.
\input{Egbert2010}

\bibitem{Blackburn1998}
Blackburn N, Fenchel T, Mitchell J (1998) {Microscale Nutrient Patches in
  Planktonic Habitats Shown by Chemotactic Bacteria}.
\newblock Science 282: 2254--2256.
\input{Blackburn1998}

\bibitem{Stocker2012}
Stocker R, Seymour JR (2012) {Ecology and physics of bacterial chemotaxis in
  the ocean}.
\newblock Microbiology and Molecular Biology Reviews 76: 792--812.
\input{Stocker2012}

\bibitem{Stocker2008}
Stocker R, Seymour JR, Samadani A, Hunt DE, Polz MF (2008) {Rapid chemotactic
  response enables marine bacteria to exploit ephemeral microscale nutrient
  patches}.
\newblock Proceedings of the National Academy of Sciences of the United States
  of America 105: 4209--14.
\input{Stocker2008}

\bibitem{Johansen2002}
Johansen JE, Pinhassi J, Blackburn N, Zweifel UL, Hagstrom A (2002)
  {Variability in motility characteristics among marine bacteria}.
\newblock Aquatic Microbial Ecology 28: 229--237.
\input{Johansen2002}

\bibitem{Ahmed2010}
Ahmed T, Shimizu TS, Stocker R (2010) {Microfluidics for bacterial chemotaxis}.
\newblock Integrative Biology 2: 604--29.
\input{Ahmed2010}

\bibitem{Blackburn1999}
Blackburn N, Fenchel T (1999) {Influence of bacteria, diffusion and shear on
  micro-scale nutrient patches, and implications for bacterial chemotaxis}.
\newblock Marine Ecology Progress Series 189: 1--7.
\input{Blackburn1999}

\bibitem{Celani2010}
Celani A, Vergassola M (2010) {Bacterial strategies for chemotaxis response}.
\newblock Proceedings of the National Academy of Sciences of the United States
  of America 107: 1391--6.
\input{Celani2010}

\bibitem{Clark2005}
Clark DA, Grant LC (2005) {The bacterial chemotactic response reflects a
  compromise between transient and steady-state behavior}.
\newblock Proceedings of the National Academy of Sciences of the United States
  of America 102: 9150--5.
\input{Clark2005}

\bibitem{Matlab}
{The MathWorks, Inc, Natick, Massachusetts, United States} (2013).
\newblock {MATLAB and Image Processing Toolbox Release 2013a}.
\input{Matlab}

\bibitem{Berg1972}
Berg HC, Brown DA (1972) {Chemotaxis in Escherichia coli analysed by
  Three-dimensional Tracking}.
\newblock Nature 239: 500--504.
\input{Berg1972}

\bibitem{Segall1986}
Segall JE, Block SM, Berg HC (1986) {Temporal comparisons in bacterial
  chemotaxis}.
\newblock Proceedings of the National Academy of Sciences of the United States
  of America 83: 8987--91.
\input{Segall1986}

\bibitem{Sourjik2002}
Sourjik V, Berg HC (2002) {Receptor sensitivity in bacterial chemotaxis}.
\newblock Proceedings of the National Academy of Sciences of the United States
  of America 99: 123--7.
\input{Sourjik2002}

\bibitem{Mitchell2002}
Mitchell JG (2002) {The energetics and scaling of search strategies in
  bacteria}.
\newblock The American Naturalist 160: 727--40.
\input{Mitchell2002}

\bibitem{Schnitzer1993}
Schnitzer MJ (1993) {Theory of continuum random walks and application to
  chemotaxis}.
\newblock Physical Review E 48: 2553--68.
\input{Schnitzer1993}

\bibitem{DeGennes2004}
de~Gennes PG (2004) {Chemotaxis: the role of internal delays}.
\newblock European Biophysics Journal 33: 691--3.
\input{DeGennes2004}

\bibitem{Gotz1987}
Gotz R, Schmitt R (1987) {Rhizobium meliloti swims by unidirectional,
  intermittent rotation of right-handed flagellar helices}.
\newblock Journal of Bacteriology 169: 3146--50.
\input{Gotz1987}

\bibitem{Armitage1987}
Armitage JP, Macnab RM (1987) {Unidirectional, intermittent rotation of the
  flagellum of Rhodobacter sphaeroides}.
\newblock Journal of Bacteriology 169: 514--8.
\input{Armitage1987}

\bibitem{Kafri2008}
Kafri Y, da~Silveira RA (2008) {Steady-state chemotaxis in Escherichia coli}.
\newblock Physical Review Letters 100: 1--4.
\input{Kafri2008}

\end{thebibliography}


\section*{References}



\end{document}